\documentclass[12pt]{article}
\usepackage{amssymb,amsmath}
\usepackage{xy} 
\xyoption{all} 

\newcommand{\sect}[1]{\setcounter{equation}{0}\section{#1}}

\title{\vspace{-1in}
\parbox{\linewidth}
{\small\hfill math-ph/0208008}\\
\vspace{0.6in}
{\bf Geometric Quantization}}
\author{\textsc{William Gordon Ritter}\thanks{email:
ritter@fas.harvard.edu}\\
\emph{Jefferson Physical Laboratory, Harvard University}\\
\emph{Cambridge, MA 02138, USA}} 

\begin{document}
\pagestyle{plain}
\setcounter{page}{1}
\newcounter{bean}
\baselineskip16pt

\maketitle

\begin{abstract}

We review the definition of geometric quantization, which begins with
defining a mathematical framework for the algebra of observables that holds
equally well for classical and quantum mechanics. We then discuss
prequantization, and go into details of the general method of
quantization with respect to a polarization using densities and
half-forms. This has applications to the theory of unitary group
representations and coadjoint orbits. 

\end{abstract}

\vspace{0.5in}

\tableofcontents

\def\H{\mathcal{H}}
\def\O{\mathcal{O}}
\def\P{\mathcal{P}}
\def\A{\mathfrak{A}}
\def\g{\mathfrak{g}}

\def\Ham{\operatorname{Ham}}
\def\Tr{\operatorname{Tr}}
\def\Symp{\operatorname{Symp}}
\def\image{\operatorname{Im}}
\def\id{\operatorname{id}}
\def\Lin{\operatorname{Lin}}
\def\supp{\operatorname{supp}}

\def\R{\mathbb{R}}
\def\Z{\mathbb{Z}}
\def\C{\mathbb{C}}
\def\B{{\cal B}}

\def\del{\nabla} 
\def\To{\longrightarrow} 
\def\d{\partial} 
\def\eps{\epsilon} 
\def\({\left(} 
\def\){\right)} 
\def\<{\left\langle} 
\def\>{\right\rangle} 
\def\Mapsto{\longmapsto} 
\def\eqdef{\overset{\te{def}}{=}}
\def\fZ{{\frak Z}}
\def\half{\frac{1}{2}}

\def\te#1{\text{#1}}
\def\t#1{\widetilde{#1}}
\def\tilde#1{\widetilde{#1}}
\def\hat#1{\widehat{#1}}
\def\norm#1{\left\| #1 \right\|} 
\def\prooflem#1{\vskip 0.05 in 
        \noindent {\it Proof of Lemma \ref{#1}.} $\ \ $ } 
\def\proofthm#1{\vskip 0.05 in 
        \noindent {\it Proof of Theorem \ref{#1}.} $\ \ $ } 
\def\endproof{$\ \ /\!\!/$ \vskip 0.05 in} 
\def\at#1#2{\left. #1 \right|_{#2}} 
\def\derpar#1#2{\frac{\partial{#1}}{\partial{#2}}}

\newtheorem{definition}{Definition} 
\newtheorem{conjecture}{Conjecture} 
\newtheorem{theorem}{Theorem} 
\newtheorem{lemma}{Lemma} 

\newpage
\sect{Introduction} 

The basic problem of quantization is the relationship between
observables of classical systems and quantum systems. It is also 
an opportunity for a bridge to be built between 
mathematics and physics, since the problem of quantization is
motivated by physical 
concerns, but the technical difficulties involve sophisticated
mathematics. Quantum mechanical states are represented by rays in a
Hilbert space $\H$, and the 
observables are represented by symmetric operators on $\H$. In
classical mechanics the state space is a symplectic manifold $(M, \omega)$ 
and observables are smooth functions, i.e. elements of $C^\infty(M, \R)$. 

Taking the view in quantum mechanics that the observables evolve in
time while the states 
remain fixed is known in physics as the Heisenberg picture. The
fundamental equation describing the dynamical evolution of a
particular (time-dependent) observable $A_t$ is the famous 
\emph{Heisenberg equation} 
$\frac{dA_t}{dt} = -\frac{i}{\hbar} [H, A_t]$, where $H$ is the 
energy observable. This is directly analogous to the situation in
classical mechanics. If $(M,\omega)$ is the symplectic phase space of
a classical system, then the dynamics of a 
time-evolving observable $f_t : M \times \R \To \R$ is
given by the differential equation 
\begin{equation} 
\frac{\d f_t}{\d t} = \{H, f_t\}, 
\label{hameqn} 
\end{equation} 
where $\{ \, , \, \}$ denotes the Poisson bracket. For the canonical 
choice of symplectic structure on $T^*\R^n$, Eq.~\eqref{hameqn} is equivalent to 
Hamilton's equations of motion as presented in \cite{goldstein}. 

The starting point of geometric quantization is to hope that the 
relationship between Heisenberg's equation and Hamilton's equation 
exhibited above is a special case of some 
general situation of deeper mathematical meaning. 

\sect{The Mathematical Structure of Physics} 

In this section we describe a general mathematical framework for
physical theories.Classical mechanics and quantum mechanics are both
realizations of this framework; thus, 
it is an important starting point for quantization. This was inspired in part by lectures 
of L.~Faddeev \cite{Deligne:qp}. The fundamental objects are a set
$\A$ of observables, a 
set $\Omega$ of states, and a \emph{probability interpretation map} 
$\A \times \Omega \to \P$, where $\P$ denotes the set of all nonnegative 
Lebesgue measurable functions $f : \R \to \R$ such that 
$\int_{-\infty}^\infty f(x) \, dx = 1$ (i.e. probability
distributions). For a state $\eta$ and an observable $A$, we write 
the associated probability distribution function as 
$\eta_A(\lambda)$. Of course, there is a natural mean-value map from
$\P \To \R$, given by $f \Mapsto \int \lambda f(\lambda) \, 
d\lambda$. In all useful examples, $\A$ and $\Omega$ both have the
structure of real vector spaces, and the composition 
\[
\A \times \Omega \To \P \overset{\te{mean-value}}{\To} \R, \qquad \eta,\, A \Mapsto 
\<\eta | A \>
\]
defines a duality between states and observables. It is clear in
physics that certain observables are not independent but rather they
are mathematical functions of other, more fundamental
observables. An example is the observable $p^2$ for a 
classical harmonic oscillator, where $p$ denotes the 
momentum vector. This fits into the framework above as follows. Given 
a real function $f : \R \to \R$ and observables $A,B$, we write $B =
f(A)$ provided that 
\[
\< \eta | B \> = \int f(\lambda) d\eta_A(\lambda) \ \te{ for all states } \ \eta
\]
In all known cases of practical importance, $\A$ has the structure of
an algebra, and in case $f(x) = \sum \alpha_i x^i$ is a polynomial
function, we have $f(A)$, as defined above, equal to $\sum \alpha_i A^i$. 
Finally, one takes as part of the data a Lie bracket $\{\, , \, \}$ on $\A$ which is an 
algebra derivation. A fixed observable $H$, the \emph{Hamiltonian}, 
is chosen on physical grounds; $H$ equals the total energy and is 
such that the differential equation $\frac{dA_t}{dt} = \{H, A_t\}$
generates the correct dynamical evolution of observables.

One can reconstruct all features of classical mechanics (even classical 
statistical mechanics) with the additional assumption that the algebra $\A$ 
is commutative. In this situation, there exists a symplectic manifold 
$(M, \omega)$ s.t. $\A = C^\infty(M)$ as algebras, in which case
we take states as normalized 
measures $\eta$ on $M$ (a statistical mechanics description), 
the probability interpretation map as 
\[
\eta, \, f \Mapsto \eta_f(\lambda) \eqdef \int_M \theta(\lambda - f(m)) \, 
d\eta(m), \qquad \theta = \te{ step function},   
\]
and $\{f , g \} = \omega(X_f, X_g)$ as the Poisson bracket.
Nonstatistical mechanics falls out of this by considering a 
restricted state space (called \emph{pure states}): the space of atomic 
measures concentrated at a single point of $M$, which is of course 
naturally identified with $M$ itself. 

In the case of quantum mechanics, the algebra of observables $\A$ is
usually realized as an algebra of linear operators on a complex
Hilbert space $\H$, and the space $D(\H)$ of positive 
operators with unit trace (or \emph{density matrices}) is taken as the
space of states. In particular, this state space contains 
the projective Hilbert space (pure states)
\[
\mathbb{P}(\H) = \{\te{all projection operators onto 1-dimensional subspaces}\} 
\ \cong \ \H/\sim
\] 
where $\sim$ is equivalence modulo multiplication by a phase. 
Elements of $\mathbb{P}(\H)$ are known as pure states, while elements
of $D(\H)$ which cannot be represented as one-dimensional projectors
are known as mixed states. The probability interpretation between 
a state $\eta$ and an observable $A$ is given by the pairing 
\[
\eta_A(\lambda) = \Tr_\H(\eta P_A(\lambda)) 
\]
where $P_A(\lambda)$ is the projector function associated to the operator
$A$ by the spectral theorem. The dynamical bracket is $\{A,B\} =
(i/\hbar)(AB - BA)$, which completes the specification of quantum 
mechanics in terms of the structure above.

\sect{Prequantization} \label{preq} 

Let $(M, \omega)$ be a $2n$-dimensional symplectic manifold and let 
\[
\eps = \( \frac{1}{2\pi \hbar} \)^n dp_1 \wedge dp_2 \wedge \dots \wedge dp_n 
\wedge dq^1 \wedge dq^2 \wedge \dots \wedge dq^n
\]
be the natural volume element. 
Based ultimately on physical experiment, Dirac formulated the
following 
prescriptions of the mathematical structure of quantization around
1925, 
long before mathematicians knew that the procedure was possible. A 
suitable quantization will produce a quantum mechanical Hilbert space 
$\H$ from $M$ in a natural way, and will associate to each classical 
observable $f : M \To \R$ an operator $\hat f$, possibly unbounded, 
acting on $\H$. On physical grounds, the mapping $C^\infty(M) \To \O$ 
given by $f \Mapsto \hat f$ should at least satisfy the following properties: 
\begin{enumerate} 
\item[(Q1)] $f \Mapsto \hat f$ is $\R$-linear. 
\item[(Q2)] if $f$ is constant, i.e. $f(m) = \alpha$ for all $m \in M$ and for some fixed real number $\alpha$, then $\hat f = \alpha I$, where $I$ is the identity operator on $\H$. 
\item[(Q3)] if $\{f_1, f_2\} = f_3$, then $[\hat f_1, \hat f_2] = -i \hbar \hat f_3$.
\end{enumerate} 
If the hat operation is to be a bijective correspondence 
$C^\infty(M) \overset{\text{1-1,onto}}{\longleftrightarrow} \O$, then the Hilbert space 
needed is too large to be physically meaningful. However, choosing a polarization of $M$ 
determines a subalgebra of classical observables which can 
admit bijective quantization maps satisfying Q1-Q3, with the added bonus that the 
associated Hilbert space is also the space of states of a known quantum mechanical system. 
We will return to this point. 

We also require the \emph{irreducibility postulate}: 
if $\{ f_j \}$ is a complete set of classical observables
of $(M,\Omega )$, then the Hilbert space $\H$ has to be irreducible
under the action of the set $\{ \hat f_j \}$.
Alternatively, suppose $G$ is a group of symmetries of a physical system
both for the classical and quantum descriptions.
If $G$ acts transitively on $(M, \Omega )$,
then $\H$ is an irreducible representation space
for a $U(1)$-central extension of the corresponding
group of unitary transformations.

Since any symplectic manifold will have a natural volume element
$\eps$, and hence a 
natural measure $d\eps$, there will also be a natural Hilbert space
$\H = L^2(M, d\eps)$. 
Each $f \in C^\infty(M)$ acts on $\H$ by a symmetric operator $-i
\hbar X_f$, and this 
correspondence satisfies Q1 and Q3, but not Q2. However, by modifying
this definition 
appropriately and using a little gauge theory, one arrives at the
construction 
known as \emph{prequantization}, which we describe presently. 

\begin{definition} \label{IC} 
A symplectic manifold $(M, \omega)$ is said to be \emph{quantizable}
if $\omega$ satisfies 
the integrality condition, i.e. if the class of $(2\pi \hbar)^{-1}
\omega$ in 
$H^2(M,\R)$ lies in the image of $H^2(M, \Z)$. 
\end{definition} 

The integrality condition which appears in Definition \ref{IC} is
equivalent to the 
statement that there exists a Hermitian line bundle $B \To M$ and a
connection $\nabla$ on 
$B$ with curvature $\hbar^{-1} \omega$. It is this latter form of the
integrality 
condition (IC) which we will actually use. In this situation, the
space of inequivalent 
pairs $(B, \del)$ is parametrized by $H^1(M, S^1)$. This is
significant because if $M$ is 
simply connected, then $H^1(M, S^1)$ is trivial and there is a unique
choice of $B$ and 
$\del$. A bundle $B \To M$ with connection chosen as above is called a
\emph{prequantum 
bundle}. Let $(\, , \, )$ be the Hermitian structure on the bundle
$B$, and let $\H = 
L^2(M, B)$, the space of square-integrable sections of the prequantum
bundle, with the 
inner product $\< s, s'\> = \int_M (s, s') \eps$. 
For $f \in C^\infty(M)$, define a symmetric operator $\hat f$ 
initially on smooth sections of $\H$ by 
\[
\hat f \, s = -i \hbar \del_{X_f} s + f\, s
\]
This choice of $\H$ and of the map $f \Mapsto \hat f$ satisfies Q1-Q3, but the Hilbert 
space constructed is too large to represent the space of states 
of any physically reasonable quantum system. For a function $f$ on $M$
such that the Hamiltonian vector field $X_f$ is complete, the
one-parameter group ${\phi_f}^t$ of canonical transformations
generated by $f$ preserves the scalar product  $\< s, s'\> = 
\int_M (s, s') \eps$, and therefore $\hat f$ extends to a self-adjoint
operator on $\H$. However, if we wish to give a probabilistic
interpretation to the scalar product by associating to $\< \lambda,
\lambda \>(x)$ the probability density of finding the quantum state
described by $\lambda$ in the classical state described by the point
$x$ in classical phase space, we would violate the uncertainty
principle since square-integrable sections of $B$ can have arbitrarily
small support. Intuitively, a position-space or momentum-space representation
corresponds to a certain choice of polarization. Without
introducing polarizations, it is no longer true that a wave function
sharply peaked in the position variables cannot also be sharply peaked
in the momentum variables. 

For this reason, the construction outlined above is called 
\emph{prequantization}, and a refinement of some sort is needed before
this procedure can rightly be called ``quantization.''

\sect{Quantization}

In quantum mechanics one may represent the Hilbert space as the space
of square-integrable complex functions on the spectrum of any complete
set of commuting observables. A natural classical analogue of a
complete set of commuting observables is a collection of $n = \half
\dim M$ functions $f_1, \ldots, f_n$ on $M$, independent at all points
of $M$ where they are defined, such that the Hamiltonian vector
fields $X_{f_i}$ are complete, and such that $\{f_i, f_j\} = 0$ for
all $1 \leq i,j \leq n$. The Hamiltonian vector fields $X_{f_i}$ span
over $\C$ an involutive distribution $F$ such that (i) $\dim_\C(F) =
\half \dim(M)$ and (ii) $\at{\omega}{F \times F} = 0$. A complex
distribution $F$ satisfying (i)-(ii) is called a \emph{Lagrangian
distribution}. A \emph{polarization} is a complex involutive
Lagrangian distribution $F$ such that $\dim(F_x \cap \bar{F}_x)$ is
constant over $x \in M$. The complex distributions $F \cap \bar{F}$
and $F + \bar{F}$ are complexifications of certain real distributions
traditionally denoted $D$ and $E$ in the literature. The vector spaces
given by $D$ and $E$ at a point are $\omega$-perpendicular. The
involutivity of $F$ implies that $D$ is involutive, so $D$ is a
foliation. We let $\pi_D : M \to M/D$ denote the projection onto the
space of integral manifolds. A polarization $F$ is said to be
\emph{admissible} if $E$ is also a foliation, the spaces $M/D$ and
$M/E$ are quotient manifolds of $M$, and the canonical projection
$\pi_{DE} : M/D \to M/E$ is a submersion. For admissible
polarizations, the tangent bundle $T\Lambda$ of each integral manifold
$\Lambda$ of $D$ is globally spanned by commuting vector fields. Also,
a Hamiltonian vector field $X_f$ lies entirely in $D$ if and only if
$f$ is constant along $E$. Furthermore, each fiber $N$ of $\pi_{DE}$
has a K\"ahler structure such that $\at{F}{\pi_D^{-1}(N)}$ projects
onto the distribution of anti-holomorphic vectors on $N$. For an
admissible positive (this means $i\omega(\xi, \bar{\xi}) \geq 0$ for
all $\xi \in F$)  polarization, the K\"ahler metric on $N$ is positive
definite. A polarization $F$ is said to be \emph{real} if $F =
\bar{F}$. Any real-polarized symplectic manifold is locally
symplectomorphic to a cotangent bundle with its vertical polarization.

Given a polarization $F$ of $(M, \omega)$ with a prequantum line
bundle $B$, one could take sections of $B$ which are covariantly
constant along $F$ to form the representation space, except that if
$\lambda_1, \lambda_2$ are two such sections, then $\<\lambda_1,
\lambda_2\>$ is constant along $D$ and hence 
$\int_M \<\lambda_1,\lambda_2\> \eps$ diverges generically unless the
leaves of $D$ are compact. Since $\<\lambda_1,\lambda_2\>$ defines a
function on $M/D$, one could define a scalar product by integrating
$\<\lambda_1,\lambda_2\>$ over $M/D$, except that there is no
canonically defined measure on $M/D$. The strategy is then to tensor
$B$ with another bundle so that $\<\lambda_1,\lambda_2\>$ may be
promoted to a \emph{density} on $M/D$ rather than a scalar
function. Tensoring $B$ with $\sqrt{\wedge^n F}$, as we shall see,
will lead to the correct modification of the Bohr-Sommerfeld
conditions, and will enable one to construct unitary representations
of certain groups of canonical transformations.

The collection of all linear frames of $F$ forms a principal
$GL(n,\C)$ fibre bundle $\B F$ over $M$, and associated to this frame
bundle is the complex line bundle $\wedge^n F$. Let 
$ML(n,\C) \overset{\rho}{\To} GL(n,\C)$ denote the double covering
group of $GL(n,\C)$. A \emph{bundle of metalinear frames} of $F$ is a
right principal $ML(n,\C)$ fibre bundle $\tilde{\B}F$ over $M$,
together with a map $\tau : \tilde{\B}F \to \B F$ such that the following
diagram commutes: 
\[
\xymatrix{ 
\tilde{\B}F \times ML(n,\C) \ar[r] \ar[d]^{\tau \times \rho} &
\tilde{\B}F \ar[d]^{\tau}  \\ 
\B F \times GL(n,\C) \ar[r] & \B F 
}
\]
where the horizontal arrows denote group actions. Let $\chi : ML(n,\C)
\to \C$ denote the unique holomorphic square root of the complex
character $\det \circ \rho$ of $ML(n,\C)$ such that $\chi(I) = 1$. 
We define $\sqrt{\bigwedge^n F}$ to be the fibre bundle associated to
$\tilde{\B}F$ with standard fibre $\C$ on which a typical element $C
\in ML(n,\C)$ acts by multiplication by $\chi(C)$. 
The space of sections $\mu$ of $\bigwedge^n F$ is isomorphic to the
space of complex valued functions on $\B F$ satisfying 
$\mu^{\#}(w C) = \det(C^{-1}) \mu^{\#}(w)$ for all $w \in \B_x F$ and
$C \in GL(n,\C)$, with the isomorphism being
$\mu^{\#} \Mapsto \mu(w) \equiv \mu^{\#}(w_1, \ldots, w_n) w_1 \wedge
\dots \wedge w_n$. Similarly, the space of
sections of $\sqrt{\bigwedge^n F}$ is isomorphic to the space of
functions $\nu^\#$ on $\tilde{\B}F$ satisfying $\nu^\#(\tilde{w}C) =
\chi(C^{-1}) \nu^\#(\tilde{w})$ for $\tilde{w} \in \tilde{\B}F$ and $C
\in ML(n,\C)$.

Quantum states of the system under consideration are represented by
sections of $B \otimes \sqrt{\bigwedge^n F}$ which are covariantly
constant along $F$. If $\sigma$ is suc a section, and if $\psi$ is a
complex-valued function on $M/D$ which is holomorphic when restricted
to fibres of $\pi_{DE}$, then $(\psi \circ \pi_D)\sigma$ is also a
section of $B \otimes \sqrt{\bigwedge^n F}$ covariantly
constant along $F$. Thus quantum states may be represented by sections
of $B \otimes \sqrt{\bigwedge^n F}$ which are covariantly
constant along $D$ and holomorphic along fibres of $\pi_{DE}$. 

To each pair $(\sigma_1, \sigma_2)$ of sections of $B \otimes
\sqrt{\bigwedge^n F}$ covariantly constant along $F$, we associate a
complex density $\< \sigma_1, \sigma_2 \>_{M/D}$ on $M/D$. For each $x
\in M$, there is a neighborhood $V \ni x$ such that 
\[
\at{\sigma_i}{V} = \lambda_i \otimes \nu_i \quad (i=1,2) 
\]
where $\lambda_i$ are covariantly constant sections of $\at{B}{V}$ and
$\nu_i$ are covariantly constant sections of $\at{\sqrt{\bigwedge^n F}}{V}$.
Consider a basis
\def\u{\overline{u}} 
\def\b{\underline{b}}
\begin{equation} \label{basis} 
(v_1, \ldots, v_d, u_1, \ldots, u_{n-d}, \u_1, \ldots, \u_{n-d}, w_1,
\ldots, w_d) 
\end{equation} 
of $T_x^\C M$ such that $\{v_i\}$ is a basis of $D_x$, $\b = (v_1,
\ldots, v_d, u_1, \ldots, u_{n-d})$ is a basis of $F_x$, and for $1
\leq i,j \leq d$ and $1 \leq k,r \leq n-d$ we have 
\[
\omega(v_i, w_j) = \delta_{ij}, \quad 
i \omega(u_k, \u_r) = \delta_{kr} \quad 
\omega(u_k, w_j) = \omega(w_i, w_j) = 0 
\]
The basis \eqref{basis} projects under $\pi_D$ to a basis $\xi_{x,D}$ of
$T^\C_{\pi_D(x)}M/D$. The value of 
\begin{equation} \label{density} 
\<\lambda_1(x), \lambda_2(x)\> \nu_1^{\#}(\t{\b}) 
\overline{\nu_2^{\#}(\t{\b})} 
\end{equation} 
(where $\t{\b}$ is a metalinear frame of $F$ at $x$ projecting onto
$\b$) depends only on $\sigma_1, \sigma_2$ and the projected basis
$\xi_{x,D}$ of
$T^\C_{\pi_D(x)}M/D$, hence we define \eqref{density} to be the value of
the density $\< \sigma_1, \sigma_2 \>_{M/D}$ on the basis
$\xi_{x,D}$. Hence the sesquilinear form 
\[
( \sigma_1 \mid \sigma_2 )_c := \int_{M/D} \< \sigma_1, \sigma_2
\>_{M/D}
\]
is a Hermitian inner product on the Hilbert space $\H^0$ defined as
the completion of the pre-Hilbert space of sections $\sigma$ such that
$(\sigma \mid \sigma)_c < \infty$. Note that $\H^0$ is the subspace of
the full representation space $\H$ corresponding to the continuous
spectrum of the complete set of commuting observables used to define
the representation. If the polarization is real and the integral
manifolds of $D$ are simply connected, then $\H^0 = \H$. 

The complement of $\H^0$ in the representation space $\H$ is spanned
by distributional sections of $B \otimes \sqrt{\bigwedge^n F}$ covariantly
constant along $F$. The supports of these sections are restricted by
Bohr-Sommerfeld conditions. Let $\Lambda$ be an integral manifold of
$D$. The operator $\del$ of covariant derivative on sections of $B
\otimes \sqrt{\bigwedge^n F}$ in the direction $F$ induces a flat
connection on $\at{(B \otimes \sqrt{\bigwedge^n F})}{\Lambda}$. Let
$G_\Lambda \subset \C^\times$ be the holonomy group of this flat
connection. The \emph{Bohr-Sommerfeld variety} is 
\[
S = \{x \in M \mid G_{\Lambda(x)} = 1 \} 
\]
where $\Lambda(x)$ is the integral manifold passing through $x$. 
Note that $S = M$ if each $\Lambda$ is simply connected. Covariantly
constant sections of $B \otimes \sqrt{\bigwedge^n F}$ vanish in $M
\setminus S$. Thus as claimed the supports of the distributional
sections are restricted by Bohr-Sommerfeld conditions. To relate this
to the classical Bohr-Sommerfeld conditions, choose a neighborhood $U$
such that $\at{B}{U}$ admits a trivializing section $\lambda$. Then
$\del \lambda = -i \hbar^{-1} \theta \otimes \lambda$ where $\theta$
is a 1-form on $U$ such that $\at{\omega}{U} = d\theta$. For each loop
$\gamma$ in $U$, the corresponding holonomy is $\exp(i\hbar^{-1}
\int_\gamma \theta)$. If $\gamma \subset \Lambda$, $\Lambda \in M/D$
then we denote by $\exp(-2\pi i d_\gamma)$ the element of the holonomy
of the flat connection on 
$\at{(B \otimes \sqrt{\bigwedge^n F})}{\Lambda}$ corresponding to
$\gamma$. The condition $G_\Lambda = 1$ is then equivalent to 
\[
\int_\gamma \theta = (n_\gamma + d_\gamma) / \hbar , \quad n_\gamma
\in \Z
\]
for each loop $\gamma$ in $\Lambda$. 

A polarization $F$ of $(M,\omega)$ is said to be \emph{complete} 
if all Hamiltonian vector fields in $F$ are complete. We will describe
the full representation space for a complete admissible real
polarization. For $k \in \{0, \ldots, n\}$ let 
\[
M_k = \{ x \in M \mid \Lambda_x \cong T^k \times \R^{n-k} \}, \quad
\text{ and } \quad S_k = S \cap M_k \, .
\]
We note that all integral manifolds of $D$ are isomorphic to products
of tori and affine spaces, so $\bigcup_{i=0}^n M_k = M$. For each $x \in
S_k$ there exists a neighborhood $V$ of $\pi_D(x)$ in $M/D$ and a
codimension $k$ submanifold $Q$ such that 
\begin{equation} \label{Q} 
\pi_D(M_k) \cap Q \subseteq \pi_D(S_k), \quad \text{ and } \quad
\pi_D(S) \cap V \subseteq Q 
\end{equation} 
Let $\Gamma_k$ denote the space of sections $\sigma$ of $B \otimes
\sqrt{\bigwedge^n F}$ satisfying (i) $\supp(\sigma) \subseteq S_k$,
(ii) $\pi_D(\supp \sigma) \subset M/D$ is compact, and (iii) for each
$x \in S_k$, $\at{\sigma}{\pi_D^{-1}(Q)}$ is a smooth section of 
$\at{B \otimes \sqrt{\bigwedge^n F}}{\pi_D^{-1}(Q)}$ covariantly
constant along $\at{F}{\pi_D^{-1}(Q)}$. Let $\sigma_1, \sigma_2 \in
\Gamma_k$. By condition (ii), there exist a finite number of disjoint
submanifolds $Q_1, \ldots, Q_s$ of $M/D$ satisfying \eqref{Q} such
that $\supp(\sigma_j) \subseteq \bigcup_{i=1}^s \pi_D^{-1}(Q_i)$. The
pair $\sigma_1, \sigma_2$ defines on $Q_i$ a density $\<\sigma_1,
\sigma_2\>_{Q_i}$ by the same
procedure as for the continuous part of the spectrum discussed
above. The scalar product on $\Gamma_k$ is defined by 
\[
\left( \sigma_1 \mid \sigma_2 \right)_k = \sum_{i=1}^s \int_{Q_i}
\<\sigma_1, \sigma_2\>_{Q_i} 
\]
and $\H^k$ is defined to be the Hilbert space completion of $\Gamma_k$
with respect to $( \ \mid \ )_k$. The full representation space is a
direct sum 
\[
\H = \bigoplus_{k=0}^n \H^k 
\]

\vskip 0.1 in \noindent {\it Quantization of Observables} 
\vskip 0.1 in 

A quantized operator is said to be \emph{polarized} with respect to a
particular polarization $P$ iff it maps polarized sections to other
polarized sections. It is necessary to work with polarized operators
in order to satisfy the irreducibility postulate. Therefore, we
briefly discuss necessary and sufficient conditions for polarized
operators. We let $\H_P$ denote a Hilbert space of $P$-polarized sections
coming from a prequantization bundle. 

Now $\hat f$ 
maps $\H_P \to \H_P$ only if the flow of $X_f$ preserves $P$. In 
particular, if $\hat{f}\H_P \subset \H_P$ then $[X, X_f] \in 
V_P(M)$ whenever $X \in V_P(M)$, so only a limited class of observables can 
be quantized. The elements of $C^\infty(M)$ that can be quantized are 
precisely the functions of canonical coordinates $(p,q)$ 
which can be represented locally in the form $f = v^a(q) p_a + u(q)$. 

For a given classical observable $f$, the following conditions are
equivalent: 
\begin{enumerate} 
\item  $\hat f$ is a polarized operator, 
\item  $\hat f$ preserves the polarization, in the sense that 
$[\hat f, \del_P] \psi = 0$ for every polarized section $\psi$, and  
\item $[X_f, P] \subset P$, where $X_f$ is the Hamiltonian vector
  field associated to $f$. 
\end{enumerate}

\vskip 0.1 in \noindent {\it Cotangent Bundles} 
\vskip 0.1 in 

In case $M = T^*Q$ with canonical symplectic structure $\Omega$, 
there is a natural real polarization
called the {\it vertical polarization},
which is spanned by the vector fields
$\left\{ \derpar{}{p_j} \right\}$. Taking the symplectic potential
$\theta = -p_j \d q^j$, the polarized sections are functions
$\psi \in \C (T^*Q)$ such that $\derpar{\psi}{p_j} = 0$,
i.e. those constant along the fibers of $T^*Q$,
so that $\psi = \psi (q^j)$.
The operators corresponding to the observables $q_j$ and $p_j$ 
are 
\[
{\cal O}_{q^j} = q^j \  ; \
{\cal O}_{p_j} = -i\hbar \derpar{}{q^j}
\]
This is known as the {\it Schr\"{o}dinger representation}
of $(T^*Q,\Omega )$.

Using the polarization spanned by the vector fields
$\left\{ \derpar{}{q^j} \right\}$ 
(which is also real), and taking $\theta ' = q^j \d p_j$ as symplectic
potential, $\psi = \psi (p_j)$ are the polarized sections and
the operators corresponding to $q_j$ and $p_j$ are
\[
{\cal O}_{q^j} = i\hbar \derpar{}{p_j}
\ ; \ {\cal O}_{p_j} = p_j
\]
This is the {\it momentum representation} of $(T^*Q,\Omega )$.
The relation between these representations
is the Fourier transform.

\end{document}